\documentclass[showpacs,pra,twocolumn,10pt]{revtex4}
\usepackage{graphicx}
\usepackage{amsmath,amssymb}
\usepackage{bm}

\input{tcilatex}

\begin{document}

\title{Output entanglement and squeezing of two-mode fields generated by a
single atom}
\author{Ling Zhou, Qing-Xia Mu, Zhong-Ju Liu}
\address{School of physics and optoelectronic technology, Dalian University
of Technology, Dalian 116024, P.R.China}

\begin{abstract}
A single four-level atom interacting with two-mode cavities is investigated.
Under large detuning condition, we obtain the effective Hamiltonian which is
unitary squeezing operator of two-mode fields. Employing the input-output
theory, we find that the entanglement and squeezing of the output fields can
be achieved. By analyzing the squeezing spectrum, we show that asymmetric
detuning and asymmetric atomic initial state split the squeezing spectrum
from one valley into two minimum values, and appropriate leakage of the
cavity is needed for obtaining output entangled fields.
\end{abstract}

\maketitle

\section{Introduction}

One of the most intriguing features of quantum mechanics is entanglement,
which has been recognized as a valuable resource for quantum information
process. Discontinuous variables and continuous variables entanglement, as
two kinds of entanglement resource, both have been concentrated much more
attention. Continuous variables entanglement, compared with its partner
discontinuous variables, has many advantages in quantum-information science %
\cite{1} and can be used to efficiently implement quantum information
process by utilizing the continuous quadrature variables of the quantized
electromagnetic fields.

Conventionally, two-mode squeezed state emerging from the nonlinear optical
interaction of a laser with a crystal (from parametric amplification or
oscillation ) is a typical continuous variables entanglement. Recently, it
has been shown that correlated spontaneous emission laser can also work as
continuous variables entanglement producer and amplifier [2-7]. Guzm\'{a}n %
\cite{guzman} proposed a method of generating unitary single and two-mode
field squeezing in an optical cavity with an atomic cloud. As a result of
realization of a single atom laser in experiment \cite{an,kim}, people began
to interest in generating two-mode entanglement via single-atom system
[11-16]. Morigi \cite{gm,gm2} et al have shown that a single trapped atom
allows for the generation of entangled light under certain conditions. One
of our authors Zhou \cite{zhou} has proposed generating unitary two-mode
field squeezing in a single three-level atom interacting dispersively with
two classical fields inside a doubly resonant cavity, which can produce a
macroscopic entangled light. Our group also proposed schemes to generate
continuous variables entanglement in a single atom system \cite{zh,zhou2}.
Most recently, based on the same atomic level scheme as single-atom laser
experiment [10], Kiffner [16] investigated a single atom system to generate
a two-mode entangled laser via standard linear laser theory.

Although output entanglement and squeezing have been studied extensively in
other system, the existence of a squeezing operator in the system which is
similar to that of the single atom laser experiment [10] has never been
exhibited before. In this paper, we study a similar atomic level as the
experiment in [10] ( but with two-mode fields ). However, there they studied
one mode laser, here we concentrate on the output entanglement of the
cavity. Under large detuning condition, we deduce unitary squeezing operator
of two-mode fields. By means of the input-output theory, we show that
entanglement and squeezing of two-mode fields can be achieved at the output.
This paper differ from [16] in these aspects: We use effective Hamiltonian
method to obtain a squeezing field operator decoupled from the atomic
degrees of freedom rather than by tracing the atomic degrees of freedom.
Instead of studying intracavity fields, we show output entanglement.

\section{System description and calculations}

We consider a single four-level atom trapped in a doubly resonant cavity,
see Fig. 1. The atom interacts with two nondegenerate cavity modes. The
first cavity mode couples to atomic transition $|a\rangle \leftrightarrow
|c\rangle $ with the detuning $\Delta _{1,}$ and the second mode interacts
with the atom on $|b\rangle \leftrightarrow |d\rangle $ with detuning $%
\Delta _{2\text{. }}$The two classical laser fields with Rabi frequencies $%
\Omega _{3}$ and $\Omega _{4}$ drive the transitions $|a\rangle
\leftrightarrow |d\rangle $ and $|b\rangle \leftrightarrow |c\rangle $ with
detunings $\Delta _{3}$ and $\Delta _{4},$ respectively. The atomic
configuration is the same as that in \cite{kiffner}. In the interaction
picture, the Hamiltonian is 
\begin{figure}[tp]
\includegraphics[width=7cm, height=5cm]{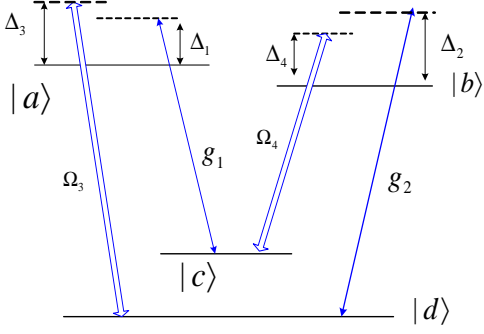}
\caption{The configuration of the atom. Two cavity modes interact with
atomic transition $|a\rangle \leftrightarrow |c\rangle $ and $|b\rangle
\leftrightarrow |d\rangle $ with detuning $\Delta _{1}$ and $\Delta _{2}$
respectively while the two classical fields $\Omega _{3}$ and $\Omega _{4}$
drive the atomic level between $|a\rangle \leftrightarrow |d\rangle $ and $%
|b\rangle \leftrightarrow |c\rangle $ with detuning $\Delta _{3}$ and $%
\Delta _{4}$, respectively.}
\end{figure}
\begin{eqnarray}
H_{1} &=&g_{1}a_{1}e^{-i\Delta _{1}t}|a\rangle \langle
c|+g_{2}a_{2}e^{-i\Delta _{2}t}|b\rangle \langle d|  \notag \\
&&+\Omega _{3}|a\rangle \langle d|e^{-i\Delta _{3}t}+\Omega _{4}|b\rangle
\langle c|e^{-i\Delta _{4}t}+h.c..
\end{eqnarray}%
Under large detuning condition $|\Delta _{k}|\gg \{|g_{j}|,|\Omega _{l}|\}$ (%
$k=1...4,j=1,2,l=3,4$), we can adiabatically eliminate the excited level $%
|a\rangle $ and $|b\rangle $ and obtain the effective Hamiltonian 
\begin{eqnarray}
H_{2} &=&(\frac{|g_{1}|^{2}}{\Delta _{1}}a_{1}^{\dagger }a_{1}+\frac{|\Omega
_{4}|^{2}}{\Delta _{4}})|c\rangle \langle c|  \notag \\
&&+(\frac{|g_{2}|^{2}}{\Delta _{2}}a_{2}^{\dagger }a_{2}+\frac{|\Omega
_{3}|^{2}}{\Delta _{3}})|d\rangle \langle d| \\
&&+[(\frac{\Omega _{3}^{\ast }g_{1}}{\Delta _{13}}a_{1}e^{i\delta _{1}t}+%
\frac{\Omega _{4}g_{2}^{\ast }}{\Delta _{24}}a_{2}^{\dagger }e^{-i\delta
_{2}t})|d\rangle \langle c|+h.c.],  \notag
\end{eqnarray}%
where $\delta _{1}=\Delta _{3}-\Delta _{1}$, $\delta _{2}=\Delta _{4}-\Delta
_{2}$, $\frac{1}{\Delta _{13}}=\frac{1}{2}(\frac{1}{\Delta _{1}}+\frac{1}{%
\Delta _{3}})$, $\frac{1}{\Delta _{24}}=\frac{1}{2}(\frac{1}{\Delta _{2}}+%
\frac{1}{\Delta _{4}})$. By making unitary transformation $U=e^{-itH_{0}}$
with 
\begin{equation}
H_{0}=\frac{|\Omega _{4}|^{2}}{\Delta _{4}}|c\rangle \langle c|+\frac{%
|\Omega _{3}|^{2}}{\Delta _{3}}|d\rangle \langle d|+\frac{\delta _{1}+\delta
_{2}}{2}(a_{1}^{\dagger }a_{1}+a_{2}^{\dagger }a_{2}),
\end{equation}%
we have the new Hamiltonian 
\begin{eqnarray}
H_{3} &=&-\frac{\delta _{1}+\delta _{2}}{2}(a_{1}^{\dagger
}a_{1}+a_{2}^{\dagger }a_{2}) \\
&&+\frac{|g_{1}|^{2}}{\Delta _{1}}a_{1}^{\dagger }a_{1}|c\rangle \langle c|+%
\frac{|g_{2}|^{2}}{\Delta _{2}}a_{2}^{\dagger }a_{2}|d\rangle \langle d| 
\notag \\
&&+[(\frac{\Omega _{3}^{\ast }g_{1}}{\Delta _{13}}a_{1}+\frac{\Omega
_{4}g_{2}^{\ast }}{\Delta _{24}}a_{2}^{\dagger })e^{i\delta t}|d\rangle
\langle c|+h.c.],  \notag
\end{eqnarray}%
where $\delta =\frac{|\Omega _{3}|^{2}}{\Delta _{3}}-\frac{|\Omega _{4}|^{2}%
}{\Delta _{4}}+\frac{\delta _{1}-\delta _{2}}{2}$. If $|\delta |\gg \{|\frac{%
\Omega _{3}^{\ast }g_{1}}{\Delta _{13}}|,|\frac{\Omega _{4}g_{2}^{\ast }}{%
\Delta _{24}}|\}$, we can perform adiabatic elimination once more and have%
\begin{eqnarray}
H_{4} &=&-\frac{\delta _{1}+\delta _{2}}{2}(a_{1}^{\dagger
}a_{1}+a_{2}^{\dagger }a_{2})  \notag \\
&&+\frac{|g_{1}|^{2}}{\Delta _{1}}a_{1}^{\dagger }a_{1}|c\rangle \langle c|+%
\frac{|g_{2}|^{2}}{\Delta _{2}}a_{2}^{\dagger }a_{2}|d\rangle \langle d| \\
&&+\frac{1}{\delta }[(\frac{\Omega _{3}^{\ast }g_{1}}{\Delta _{13}}a_{1}+%
\frac{\Omega _{4}g_{2}^{\ast }}{\Delta _{24}}a_{2}^{\dagger })(\frac{\Omega
_{3}g_{1}^{\ast }}{\Delta _{13}}a_{1}^{\dagger }+\frac{\Omega _{4}^{\ast
}g_{2}}{\Delta _{24}}a_{2})|d\rangle \langle d|  \notag \\
&&-(\frac{\Omega _{3}g_{1}^{\ast }}{\Delta _{13}}a_{1}^{\dagger }+\frac{%
\Omega _{4}^{\ast }g_{2}}{\Delta _{24}}a_{2})(\frac{\Omega _{3}^{\ast }g_{1}%
}{\Delta _{13}}a_{1}+\frac{\Omega _{4}g_{2}^{\ast }}{\Delta _{24}}%
a_{2}^{\dagger })|c\rangle \langle c|].  \notag
\end{eqnarray}%
If the atom is initially in state $|d\rangle $, we finally have the
effective Hamiltonian taken on the atomic state $|d\rangle $ as 
\begin{eqnarray}
H_{eff} &=&\lambda _{1}a_{1}^{\dagger }a_{1}+\lambda _{2}a_{2}^{\dagger
}a_{2} \\
&&+\eta a_{1}a_{2}+\eta ^{\ast }a_{1}^{\dagger }a_{2}^{\dagger },  \notag
\end{eqnarray}%
with%
\begin{eqnarray}
\lambda _{1} &=&\frac{|\Omega _{3}g_{1}|^{2}}{\delta \Delta _{13}^{2}}-\frac{%
\delta _{1}+\delta _{2}}{2}, \\
\lambda _{2} &=&\frac{|\Omega _{4}g_{2}|^{2}}{\delta \Delta _{24}^{2}}+\frac{%
|g_{2}|^{2}}{\Delta _{2}}-\frac{\delta _{1}+\delta _{2}}{2},  \notag \\
\eta  &=&\frac{g_{1}g_{2}\Omega _{3}^{\ast }\Omega _{4}^{\ast }}{\delta
\Delta _{13}\Delta _{24}}.  \notag
\end{eqnarray}%
In Eq.(6), we have thrown off a constant which does not affect the dynamics
of the system. Because the initial atomic state is $|d\rangle $, only the
terms which take action on $|d\rangle $ survive. The stark shift $\frac{%
|g_{1}|^{2}}{\Delta _{1}}a_{1}^{\dagger }a_{1}|c\rangle \langle c|$ has no
contribution and $\frac{|g_{2}|^{2}}{\Delta _{2}}a_{2}^{\dagger
}a_{2}|d\rangle \langle d|$ remain ( see the second line in Eq.(5)). Thus, $%
\lambda _{1}$ and $\lambda _{2}$ are asymmetric in form. We will show the
effect of the asymmetry on the output squeezing and entanglement.

If the initial cavity fields are in coherent state $|\epsilon _{1},\epsilon
_{2}\rangle $ (with the help of two laser pumping, we can easy obtain the
initial two-mode \ coherent state), we can use $SU(1,1)$ algebra to obtain
evolution of wave function of the fields with $|\Psi _{f}(\tau )\rangle
=e^{-iH_{eff}\tau }|\Psi _{f}(0)\rangle $. The exact expression of the
fields evolution is a two-mode coherent-squeezed state as%
\begin{equation}
|\Psi _{f}(\tau )\rangle =S(\vartheta )|\epsilon _{1},\epsilon _{2}\rangle ,
\end{equation}%
where $\vartheta =re^{i\varepsilon }$, and the squeeze parameter $r$ ($%
\varepsilon )$ is determined by $r=\tanh ^{-1}|\tau \eta ^{\ast }b_{0}\sinh
\phi |$ (tan$\varepsilon =\func{Im}(-i\eta ^{\ast }b_{0}\sinh \phi )/\func{Re%
}(-i\eta ^{\ast }b_{0}\sinh \phi ))$ with $\phi ^{2}=[|\eta |^{2}-(\frac{%
\lambda _{1}+\lambda _{2}}{2})^{2}]\tau ^{2}$, $b_{0}=[\phi \cosh \phi
+i\tau (\lambda _{1}+\lambda _{2})/2\sinh \phi ]^{-1}$ \cite{zhou}. The
evolution time $\tau $ is limited by the $\tau _{diss}=\min (\frac{1}{\kappa
_{1}}$, $\frac{1}{\kappa _{2}}$ ) where $\kappa _{1}$ and $\kappa _{2}$ are
the decay rates of modes 1 and 2. So, the intensity of the fields can not be
increased into infinity with time evolution although the initial coherent
state can effectively enhance the intensity of the cavity fields. Actually,
the intensity of fields can not be increased largely due to the loss of the
cavity and the large detunings condition. Consequently, the adiabatic
elimination still can be used within $\tau _{diss}$ only if the intensity of
the quantum fields is not larger than the intensity of two classical fields $%
\Omega _{3}$ and $\Omega _{4}$. The decay effects will be discussed in next
section where we do not need narrow the evolution time because physical
quantities are automatically limited by time evolution after considering the
decays. On the other hand, from Eq.(7) we see that the enhanced intensity of
the fields do not affect the entanglement of the two modes because the
entanglement results from the squeeze parameter.

\section{Output squeezing and entanglement}

We now concentrate on the squeezing properties of the outgoing cavity fields
which can be detected and used as entanglement source. To evaluate the
entangled light outside the cavity, we employ the input-output theory \cite%
{noise, wall}. We assume that the two cavity modes are driven by external
laser fields besides the interaction with the atom in Eq.(6). The classical
laser drive the cavity modes with strengths $\mu _{1}$ and $\mu _{2}$,
respectively. The Langvein equations of motion for the two-mode fields are
given by

\begin{eqnarray}
\dot{a}_{1} &=&-i\lambda _{1}a_{1}-i\mu _{1}^{\ast }-i\eta ^{\ast
}a_{2}^{\dagger }-\frac{\kappa _{1}}{2}a_{1}-\sqrt{\kappa _{1}}a_{1in}, \\
\dot{a}_{2} &=&-i\lambda _{2}a_{2}-i\mu _{2}^{\ast }-i\eta ^{\ast
}a_{1}^{\dagger }-\frac{\kappa _{2}}{2}a_{2}-\sqrt{\kappa _{2}}a_{2in}. 
\notag
\end{eqnarray}%
Here, $a_{1in}$ and $a_{2in}$ are annihilation operators associated with the
input fields, and $\kappa _{1}$ and $\kappa _{2}$ are the cavity decay rates
of modes $a_{1}$ and $a_{2}$. Using the transformation%
\begin{eqnarray}
a_{1} &=&a_{1}^{^{\prime }}+\alpha _{0}, \\
a_{2} &=&a_{2}^{^{\prime }}+\beta _{0},  \notag
\end{eqnarray}%
we can rewrite the Eq. (9) as

\begin{eqnarray}
\dot{a}_{1}^{^{\prime }} &=&-i\lambda _{1}a_{1}^{^{\prime }}-i\eta ^{\ast
}a_{2}^{\prime \dagger }-\frac{\kappa _{1}}{2}a_{1}^{^{\prime }}-\sqrt{%
\kappa _{1}}a_{1in}, \\
\dot{a}_{2}^{^{\prime }} &=&-i\lambda _{2}a_{2}^{^{\prime }}-i\eta ^{\ast
}a_{1}^{^{\prime }\dagger }-\frac{\kappa _{2}}{2}a_{2}^{^{\prime }}-\sqrt{%
\kappa _{2}}a_{2in},  \notag
\end{eqnarray}%
where $\alpha _{0}=\frac{-2i\mu _{1}^{\ast }(\kappa _{2}+2i\lambda
_{2})-4\mu _{2}^{\ast }\eta ^{\ast }}{(\kappa _{1}+2i\lambda _{1})(\kappa
_{2}+2i\lambda _{2})+4\eta ^{\ast 2}}$, $\beta _{0}=\frac{-2i\mu _{2}^{\ast
}(\kappa _{1}+2i\lambda _{1})-4\mu _{1}^{\ast }\eta ^{\ast }}{(\kappa
_{1}+2i\lambda _{1})(\kappa _{2}+2i\lambda _{2})+4\eta ^{\ast 2}}$.
Performing Fourier transformation, we can solve the above equation and then
use the relation $a_{jout}=a_{jin}+\sqrt{\kappa _{j}}a_{j}$ ($j=1,2)$ to
obtain the output fields as 
\begin{eqnarray}
a_{1out}(\omega ) &=&\sqrt{\kappa _{1}}\alpha _{0}\delta (\omega )+ \\
&&\frac{-(\alpha _{1}^{\ast }\alpha _{2}+|\eta |^{2})a_{1in}(\omega )+i\eta
^{\ast }\sqrt{\kappa _{1}\kappa _{2}}a_{2in}^{\dagger }(-\omega )}{\alpha
_{1}\alpha _{2}-|\eta |^{2}},  \notag \\
a_{2out}(\omega ) &=&\sqrt{\kappa _{2}}\beta _{0}\delta (\omega )+  \notag \\
&&\frac{i\eta ^{\ast }\sqrt{\kappa _{1}\kappa _{2}}a_{1in}^{\dagger
}(-\omega )-(\beta _{1}^{\ast }\beta _{2}+|\eta |^{2})a_{2in}(\omega )}{%
\beta _{1}\beta _{2}-|\eta |^{2}},  \notag
\end{eqnarray}%
where 
\begin{eqnarray}
\alpha _{1} &=&\frac{\kappa _{1}}{2}+i(\lambda _{1}-\omega ), \\
\alpha _{2} &=&\frac{\kappa _{2}}{2}-i(\lambda _{2}+\omega );  \notag \\
\beta _{1} &=&\frac{\kappa _{2}}{2}+i(\lambda _{2}-\omega ),  \notag \\
\beta _{2} &=&\frac{\kappa _{1}}{2}-i(\lambda _{1}+\omega ).  \notag
\end{eqnarray}%
Observing that $|\alpha _{0}|$ and $|\beta _{0}|$ are\ in proportion to $%
|\mu _{1}|$ and $|\mu _{2}|$, therefore, we see that the driving parameters $%
|\mu _{1}|$ and $|\mu _{2}|$ yield effective displacements to the two mode
output fields.

Now, we discuss the output entanglement of the fields. Define $I_{+}=\frac{1%
}{\sqrt{2}}(a_{1}+a_{1}^{\dagger }-a_{2}-a_{2}^{\dagger })$, $I_{-}=\frac{-i%
}{\sqrt{2}}(a_{1}-a_{1}^{\dagger }+a_{2}-a_{2}^{\dagger })$. The squeezing
spectrum can be defined as \cite{david}%
\begin{equation}
\langle I_{\pm }(\omega )I_{\pm }(\omega ^{\prime })+I_{\pm }(\omega
^{\prime })I_{\pm }(\omega )\rangle =2S_{\pm }(\omega )\delta (\omega
+\omega ^{\prime }),
\end{equation}%
where $I_{\pm }(\omega )$ is Fourier transformation of $I_{\pm }$. With the
definition of $I_{\pm }$, we have $S_{+}(\omega )=S_{-}(\omega )$ for
uncorrelated vacuum input noise. The squeezing spectrum has been connected
with entanglement criterion \cite{david}. The ``sum'' criterion of Duan et
al.\cite{duan} can be rewritten with $S_{\pm }(\omega )$ as 
\begin{equation}
S_{+}(\omega )+S_{-}(\omega )<2.
\end{equation}%
So, the two output modes are entangled if \cite{david} 
\begin{equation}
S_{\pm }(\omega )<1.
\end{equation}%
Thus, the time evolution of entanglement is transformed into frequency
domain. The spectrum $S_{\pm }(\omega )$ will be not only squeezing but also
entanglement judge.

We assume that the input field is in the vacuum. From Eq.(12), we have 
\begin{eqnarray}
S_{+}(\omega ) &=&\frac{||\eta |^{2}+\alpha _{2}\alpha _{1}^{\ast
}|^{2}+|\eta |^{2}\kappa _{1}\kappa _{2}}{2|\alpha _{2}\alpha _{1}-|\eta
|^{2}|^{2}} \\
&&-\frac{i\sqrt{\kappa _{1}\kappa _{2}}[\eta (|\eta |^{2}+\alpha _{1}\alpha
_{2}^{\ast })-\eta ^{\ast }(|\eta |^{2}+\alpha _{2}\alpha _{1}^{\ast })]}{%
2|\alpha _{2}\alpha _{1}-|\eta |^{2}|^{2}}  \notag \\
+\alpha _{j} &\rightarrow &\beta _{j}\text{.}  \notag
\end{eqnarray}%
\ \ \ \ \ \ \ \ \ We also find that the squeezing spectrum has not been
affected by the displacements $\alpha _{0}$ and $\beta _{0}$ because Eq.(17)
has no relation with $\alpha _{0}$ and $\beta _{0}$. $S_{+}(\omega )$ is
connected with the squeezing parameters $\eta ,$ decay rate $\kappa $, as
well as $\lambda _{1}$ and $\lambda _{2}.$%
\begin{figure}[tp]
\includegraphics[width=7cm, height=5cm]{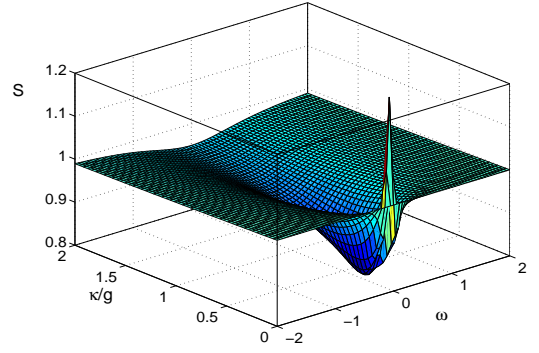}
\caption{ The output squeezing change with $\protect\omega $ and $\protect%
\kappa $. The parameters are $g_{1}=g_{2}=g,$ $\Delta _{1}=10g$, $\Delta
_{2}=12g$, $\Delta _{3}=11.5g$, $\Delta _{4}=10.5g$, $\Omega _{3}=1.5ig$, $%
\Omega _{4}=1.5g.$}
\end{figure}

Fig. 2 shows that the squeezing $S_{+}(\omega )$ change with $\kappa $ and $%
\omega $ where we choose $\kappa _{1}=\kappa _{2}=\kappa $. With the group
of the parameters, $\Delta _{k}$ is about ten times the values of \{$%
g_{j},\Omega _{l}$\}, which means that the first adiabatic elimination
condition $\Delta _{k}\gg \{|g_{j}|,|\Omega _{l}|\}$ is fulfilled. With the
parameters used in Fig. 2, we have $\delta =1.48g$, $|\frac{\Omega
_{3}^{\ast }g_{1}}{\Delta _{13}}|=0.14g$, and $|\frac{\Omega _{4}g_{2}^{\ast
}}{\Delta _{24}}|=0.13g$, so the second adiabatic elimination condition $%
|\delta |\gg \{|\frac{\Omega _{3}^{\ast }g_{1}}{\Delta _{13}}|,|\frac{\Omega
_{4}g_{2}^{\ast }}{\Delta _{24}}|\}$ satisfies. Therefore, all of the
approximation conditions are fulfilled for the parameters in Fig. 2. As
presented in Fig. 2, we see that the entanglement is achievable and $%
S_{+}(\omega )$ changes with the leakage rate. For small value of $\kappa $,
we can not obtain ideal squeezing outside the cavity. For large value of $%
\kappa $, the degree of squeezing will be decreased. That is to say, there
is a suitable value of $\kappa $ for achieving maximum degree squeezing. 
\begin{figure}[tp]
\includegraphics[width=7cm, height=5cm]{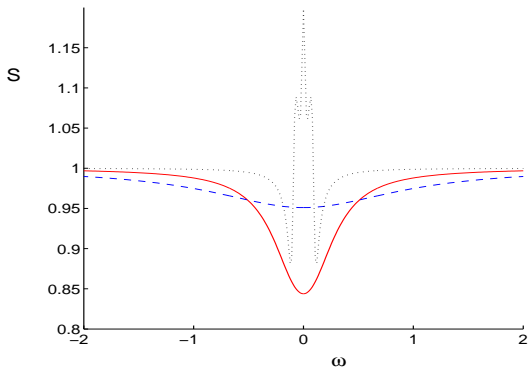}
\caption{The output squeezing spectrum for several value of loss of the
cavity where $\protect\kappa =0.05g$ (dotted line), $0.5g$ (solid line), $2g$
(dashed line). The other parameters are the same as Fig. 2.}
\end{figure}

In Fig. 3, we show the squeezing spectrum for several values of leakage rate 
$\kappa $. For $\kappa =0.05g$, we observe two minimum values in the
squeezing spectrum which also can be seen in Fig.2. However, usually the
squeezing spectrum should has one valley if $\kappa _{1}=\kappa _{2}$ \cite%
{lab} . The split from one valley into two minima is similar to the effect
of asymmetric loss for each mode \cite{wall} where if the loss of each mode
differs, the squeezing spectrum shows two minima. Here, although we set $%
\kappa _{1}=\kappa _{2}=\kappa $, we still can observe the interesting
split. Actually, the split originates from the nonzero and asymmetric $%
\lambda _{1}$ and $\lambda _{2}$ [$\lambda _{1}\neq $ $\lambda _{2}$ seen
Eq.(7)]. If $\lambda _{1}=\lambda _{2}=0$, we will have only one valley even
if for small value $\kappa $. With the increasing of $\kappa $, the split
disappears. Following the relation Eq.(13), we know that because of the
larger value of $\kappa ,$ i.e. $\kappa \gg |\lambda _{i}|$, the difference
in $\lambda _{1}$ and $\lambda _{2}$ will have little effect so that we have
one minimum squeezing. Physically, the asymmetry originates from the
asymmetric detuning and asymmetric atomic initial state. In addition, we can
also observe the existence of a appropriate value of $\kappa $ where
squeezing are better than others. For example the squeezing for $\kappa
=0.5g $ is better than that for $\kappa =0.05g,2g$. Moreover, one can see
that the bandwidth of the squeezing spectrum becomes wide when the minimum
values of squeezing are increased. 
\begin{figure}[tp]
\includegraphics[width=7cm, height=5cm]{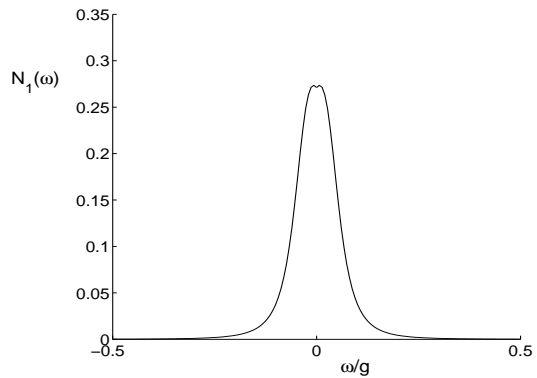}
\caption{The curve for N$_{1}$($\protect\omega $) where the parameters are $%
g_{1}=g_{2}=g,$ $\Delta _{1}=10g$, $\Delta _{2}=12g$, $\Delta _{3}=11.5g$, $%
\Delta _{4}=10.5g$, $\Omega _{3}=2.5ig$, $\Omega _{4}=2.5g$, $\protect\kappa %
_{1}=\protect\kappa _{2}=0.1g.$}
\end{figure}

We now discuss the correlation between the output fields amplitude. With
Eq.(12), we have 
\begin{eqnarray}
\langle a_{1out}^{\dagger }(\omega )a_{1out}(\omega ^{\prime })\rangle
&=&\kappa _{1}|\alpha _{0}|^{2}\delta (\omega )\delta (\omega ^{\prime }) \\
&&+\frac{|\eta |^{2}\kappa _{1}\kappa _{2}}{|\alpha _{2}\alpha _{1}-|\eta
|^{2}|^{2}}\delta (\omega -\omega ^{\prime }),  \notag \\
\langle a_{2out}^{\dagger }(\omega )a_{2out}(\omega ^{\prime })\rangle
&=&\kappa _{2}|\beta _{0}|^{2}\delta (\omega )\delta (\omega ^{\prime }) 
\notag \\
&&+\frac{|\eta |^{2}\kappa _{1}\kappa _{2}}{|\beta _{2}\beta _{1}-|\eta
|^{2}|^{2}}\delta (\omega -\omega ^{\prime })  \notag
\end{eqnarray}%
We let $N_{1}(\omega )=\frac{|\eta |^{2}\kappa _{1}\kappa _{2}}{|\alpha
_{2}\alpha _{1}-|\eta |^{2}|^{2}}$ which is one of the contributors in
intensity spectrum. In Fig.4, we plot $N_{1}(\omega )$ as a function of $%
\omega .$ We see that for resonance $\omega =0$ (the output frequency equal
to the frequency of the cavity fields ), $N_{1}(\omega )$ achieve its
maximum value but the intensity is relative smaller than that in [16]. This
might be because there [16] the detuning $\Delta _{i}$ is not so larger and
the input-output effect is not considered. However, in this paper, we
consider the atom dispersively interacts with the cavity as well as the
input-output effect, so the output fields are decreased much more. From
Eq.(17), we know that when $\omega =\omega ^{\prime }=0$, the output fields
are enhanced by $\kappa _{1}|\alpha _{0}|^{2}$ and $\kappa _{2}|\beta
_{0}|^{2}$ with $\delta $ function. With the same parameters with Fig. 4, we
have $\kappa _{1}|\alpha _{0}|^{2}\approx 19$, $\kappa _{2}|\beta
_{0}|^{2}\approx 8$ if $\mu _{1}=\mu _{2}=0.8g$. So, the fields will be
enhanced largely.

\section{Conclusion}

In summary, we study a single four-level atom interacting with two-mode
cavity system. We deduce unitary squeezing operator of the two-mode fields
via adiabatic elimination technique. By means of the input-output theory, we
show that two-mode entanglement and squeezing can be achieved at the output
fields in frequency domain. The squeezing spectrum reveals that asymmetric
detuning and asymmetric atomic initial state split the squeezing into two
minimum values, and appropriate leakage of the cavity is needed for
obtaining output entangled fields.

Acknowledgments: The project was supported by NSFC under Grant No.10774020,
and also supported by SRF for ROCS, SEM.


\begin{thebibliography}{99}
\bibitem{1} S. L. Braunstein and P. van Look, Rev. Mod. Phys. 77 (2005) 513.

\bibitem{han} H. Xiong, M. O. Scully, and M. S. Zubairy, Phys.Rev.Lett. 94
(2005) 023601.

\bibitem{tan} H. T. Tan, S. Y. Zhu,M. S. Zubairy, Phys. Rev. A 72 (2005)
022305.

\bibitem{manzoor} M. Ikram, G. X. Li and M. S. Zubairy, Phys. Rev. A 76
(2007) 042317 .

\bibitem{LGX} G. X. Li, H. T. Tan, and M. Macovei, Phys. Rev. A 76 (2007)
053827.

\bibitem{chen} Y. Ping, B. Zhang, Z. Cheng, Phys. Lett. A 366 (2007) 596.

\bibitem{zh} L. Zhou, Y. H. Ma and X. Y. Zhao, J. Phys. B 41 (2008) 215501;
Y. H. Ma, Q. X. Mu and G. H. Yang and L. Zhou, J. Phys. B, Phys. 41 (2008)
215502.

\bibitem{guzman} R. Guzm\'{a}n, J.C. Retamal, E. Solano, and N. Zagury,
Phys. Rev. Lett. 96 (2006) 010502.

\bibitem{an} K. An, J. J. Childs, R. R. Dasari, and M. S. Feld, Phys. Rev.
Lett. 73 ( 1994 ) 3375.

\bibitem{kim} J. McKeever, A. Boca, A. D. Boozer, J. R. Buck, and H. J.
Kimble, Nature (London ) 425 (2003) 268.

\bibitem{gm} G. Morigi, J. Eschner, S. Mancini, and D. Vitali, Phys. Rev.
Lett. 96 (2006) 023601.

\bibitem{gm2} G. Morigi, J. Eschner, S. Mancini, and D. Vitali, Phys. Rev. A
73 (2006) 033822.

\bibitem{zhou} L. Zhou, H. Xiong,and M. S. Zubairy, Phys. Rev. A 74, 022321
2006.

\bibitem{zhou2} Q. X. Mu, Y. H. Ma and L. Zhou, J. Phys. B 40 (2007) 3241

\bibitem{YANG} J. Jing, Z. G. L\"{u}, G. H. Yang, Physics Letters A 372
(2008) 2183

\bibitem{kiffner} M. Kiffner, M. S. Zubairy, J. Evers, and C.H. Keitel,
Phys. Rev. A 75 ( 2007 ) 033816.

\bibitem{noise} C.W. Gardiner and P. Zoller, Quantum Noise (Springer-Verlag,
Berlin, 2000).

\bibitem{wall} D. F. Walls and G. J. Milburn, \textit{Quantum Optics},
(Springer-Verlag, Berlin 1994).

\bibitem{david} D. Vitali, G. Morigi, J. Eschner, Phys. Rev. A 74 (2006)
053814 .

\bibitem{duan} L. M. Duan, G. Giedke, J. I. Cirac, and P. Zoller, Phys. Rev.
Lett. \textbf{84 }(2000) 2722.

\bibitem{lab} A. S. Labne, M. O. Reid, D. F. Wall, Phys. Rev. Lett. 60%
\textbf{\ }(1988) 1940.
\end{thebibliography}
\end{document}